# Multiply Folded Graphene

Kwanpyo Kim[1,2,3], Zonghoon Lee[4,#], Brad D. Malone[1,3], Kevin T. Chan[1,3], Benjamín Alemán[1,2,3], William Regan[1,3], Will Gannett[1,3], M. F. Crommie[1,2,3], Marvin L. Cohen[1,2,3], and A. Zettl[1,2,3]*

[1]*Department of Physics and* [2]*Center of Integrated Nanomechanical Systems, University of California at Berkeley, Berkeley, CA 94720, U.S.A.*

[3]*Materials Sciences Division, Lawrence Berkeley National Laboratory, Berkeley, CA 94720, U.S.A.*

[4]*National Center for Electron Microscopy, Lawrence Berkeley National Laboratory, Berkeley, CA 94720, U.S.A.*

[#]*Present address: School of Mechanical and Advanced Materials Engineering, UNIST (Ulsan National Institute of Science and Technology), Ulsan 689-798, S. Korea*

* To whom correspondence should be addressed: azettl@berkeley.edu

The folding of paper, hide, and woven fabric has been used for millennia to achieve enhanced articulation, curvature, and visual appeal for intrinsically flat, two-dimensional materials. For graphene, an ideal two-dimensional material, folding may transform it to complex shapes with new and distinct properties. Here, we present experimental results that folded structures in graphene, termed grafold, exist, and their formations can be controlled by introducing anisotropic surface curvature during graphene synthesis or transfer processes. Using pseudopotential-density functional theory calculations, we also show that double folding modifies the electronic band structure of graphene. Furthermore, we demonstrate the intercalation of $C_{60}$ into the grafolds. Intercalation or functionalization of the chemically reactive folds further expands grafold's mechanical, chemical, optical, and electronic diversity.

**I. Introduction**

Folding a structure changes its form and functionality. From paper origami to the coiling of proteins, folding can transform relatively simple structures into complex shapes with new and distinct physical qualities.[1-3] Graphene, an atomically thin two-dimensional layer of $sp^2$-bonded carbon atoms, has a high in-plane Young's modulus[4] but is easily warped in the out-of-plane direction, similar to a sheet of paper. Although folding induces strain energy at the curved folding edge, it can lower the total energy of the system through plane-plane interactions after folding.[5]

The folding may also induce new and distinct properties in graphene. For example, recent theoretical studies suggest that folded graphene can exhibit interesting electronic properties under magnetic fields,[6,7] such as interferometric effects due to the interplay between the gauge fields created by the fold and the external fields. Folds in the top layers of graphite[8,9] and edge-folds in suspended graphene[10-14] have been previously observed. Most of these studies have focused on the single fold structure; however, graphene folding structures can be more complex and intriguing.

In this paper, we introduce multiply folded structures in graphene, which we term grafold. Multiply folded structures, such as monolayer / triple-layer / monolayer graphene junctions, are observed in suspended and supported graphene samples and verified by various characterization techniques including transmission electron microscopy (TEM). Using pseudopotential-density functional theory calculations, we show that double-folding alters the electronic band structure of graphene. By introducing anisotropic surface curvature during the graphene synthesis or transfer processes, we also demonstrate ways to control the orientation and placement of graphene folding formations. In principle, single or periodic grafold hems, pleats, creases, ripples, and ruffles can be tailored from graphene. Furthermore, we show that grafolds can serve as a localized intercalation platform via $C_{60}$ intercalation into an edge fold. Intercalation or functionalization of the chemically reactive folds will further expand grafold's mechanical, chemical, optical, and electronic diversity.

## II. Methods

### A. Sample Preparation and Experiment

Graphene is synthesized on 25 um thick copper foil (99.8 % Alfa Aesar, Ward Hill, MA) using chemical vapor deposition (CVD).[15] After the synthesis, graphene is transferred to various substrates including Quantifoil holey carbon TEM grids.[15,16] For $C_{60}$ intercalation, graphene-covered TEM grids and ultra-pure $C_{60}$ crystals are placed in a clean quartz ampoule that is then evacuated to $10^{-6}$ Torr and sealed. The sample is then placed in a furnace at 600 °C for 7 days. After cooling, the TEM grid is lightly rinsed in isopropyl alcohol and allowed to dry.

For controlling the directions of fold formations, we synthesize graphene on a copper substrate with etched line trenches. We spincoat poly methyl-methacrylate (PMMA) on copper foil and pattern parallel lines with electron beam lithography (1 um or 2 um width, with 10 um pitch).   Then we partially etch the copper foil with $Na_2S_2O_8$ solution (concentration of 0.1 mg $Na_2S_2O_8$/1 mL water) for 1.5 minutes.   After etching, we remove the PMMA mask with acetone and grow CVD graphene on the patterned copper foil.   Alternatively, for the proposed placement control of fold formations, CVD graphene grown on an unpatterned copper foil is transferred onto a substrate which has removable patterned metal features.   We pattern various shapes into PMMA resist with electron beam lithography and thermally evaporated 3 nm Cr (sticking layer) and 150 nm Cu.   After a lift-off process, we transfer CVD graphene on top of copper structures (standard PMMA-supported transfer) and underetch copper for fold formations, followed by rinsing away etchant with deionized water and isopropyl alcohol.

For TEM imaging, we use a JEOL 2010 operated at 100 kV and the TEAM 0.5 operated at 80 kV.   TEAM 0.5, which is tuned by the image Cs aberration corrector with a monochromated beam, is used for atomic resolution imaging of graphene fold structures.   For scanning TEM (STEM) imaging and nano beam diffraction, we use a Zeiss Libra 200F operated at 200 kV.   The microscope optics use the nano parallel electron beam (NPEB) set-up optimized for fine diffraction pattern acquisition in STEM mode.   We use a configuration which gives the NPEB with diameters of 45 nm, and all diffraction patterns are energy filtered with a 40 eV width.   Scanning electron microscopy (SEM) images are aquired with an FEI XL3000.   Atomic force microscopy (AFM) images are aquired with an MFP3D scanning probe microscope (Asylum Research instruments) in non-contact mode.

## B. First Principles Calculations

First principles calculations are carried out using the pseudopotential density functional approach as implemented in the SIESTA code.[17] The local density approximation to the exchange-correlation functional is used[18,19] and the interaction of the valence electrons with the ion cores is represented by a norm-conserving pseudopotential.[20] The double-zeta atomic basis set is used with a 200 Ry cutoff for the real-space mesh. A $32 \times 1 \times 1$ k point mesh is used for the Brillouin zone integrations. Important results are crosschecked with a plane-wave basis code.

## III. Results and Discussion

The folding of graphene is not as simple as the folding of a sheet of paper, a macroscopically isotropic material without preference for folding axes. In graphene folding one must define the direction and location of folding lines in a way that accounts for the hexagonal lattice symmetry and registry (or lack thereof) between adjacent layers in the fold. As shown in Figure 1(a), the direction of folding lines (dotted blue lines), $\theta$, can be defined relative to the $\vec{a}_1 - \vec{a}_2$ direction (dotted red lines), where $\vec{a}_1$ and $\vec{a}_2$ are conventional translational unit vectors in the graphene lattice. Unless $\theta$ satisfies the equation, $\theta / 30^o = l$ where $l$ is an integer, folding will introduce a relative rotation between the resulting stacked graphene layers. Depending on the direction or location of folding line formations, certain directions of folding may be more energetically favorable than others.[8,14]

By repeating a simple folding operation in close proximity to the first fold, one can produce a variety of locally distressed and topologically altered regions, such as a monolayer / triple-layer / monolayer graphene junction.   Fig. 1a shows a planar view of graphene before double folding, where the two blue dotted lines are the symmetry axes of the fold.   An example of one possible structure is shown in Fig. 1b, where the two folding lines have different directions ($\theta_1 \neq \theta_2$).   As the two folding lines are not parallel, they will eventually converge.   Borrowing the language of the textile industry, we term this folding structure a tapered tuck.[21]

When the double folding lines are parallel ($\theta_1 = \theta_2$), the width of the trilayer region is uniform along the fold and given by $w = a\sqrt{n^2 + nm + m^2}$, where $\vec{w} = n\vec{a}_1 + m\vec{a}_2$ is the translational vector between the two folding lines (Fig. 1c).   (Note that *n* and *m* need not be integers.)   We refer to this parallel double folding structure as regular pleat folding, again referencing the analogous structure in fabrics.   In pleat folding, the top (blue) and bottom (yellow) graphene layers have the same crystalline direction, but the middle graphene layer (red) is rotated relatively to the top and bottom layers when $\theta/30^o \neq l$ (an integer).   Fig. 1d shows the top-view of a pleat folding structure where $\theta = 11^o$.   Alternatively, pleat folding formations along the zigzag or armchair directions of the graphene lattice ($\theta/30^o = l$) can produce triple graphene layers in the folding structures with no relative rotation.   However, even without relative rotation, there are several distinct stacking configurations possible, determined by the precise positions of folding line formations.   Figs. 1e-g show AAA, ABA, and ABC stacking when $\theta = 0^o$.

Using a pleat folding as a building block, we can construct many complex folding structures in graphene.  Some possible folding structures with quadruple foldings are shown in Fig. 1h.  We can also envision a superlattice of pleat foldings in graphene as illustrated in Fig. 1h and 1i.  This novel superlattice may exhibit unique electronic and optical properties distinct from planar single or multi-layer graphene with homogeneous layers, as discussed in more detail below.  Moverover, the nature of the non-homogeneous layer number and highly curved folding edges in grafold can be further manipulated.  The localized triple layer regions can serve as an ideal localized intercalation platform for encapsulating foreign atoms and molecules.  This is a new system which cannot be realized with standard graphite intercalation methods.[22]  Further, highly curved folding edges should exhibit enhanced chemical reactivity from the flat graphene regions[23] and can be used for one-dimensional functionalization of graphene.  Local grafold intercalation and edge functionalization are shown in Fig. 1i.

We now turn to the experimental observation of grafold.  Fig. 2a shows a TEM image of graphene transferred onto a Quantifoil holey carbon TEM grid.  In the middle of the suspended graphene region there appear long vertical parallel lines.  Such ribbon-shaped features are frequently seen in our CVD-grown samples.  We find that the usual widths of these features range from a few nanometers to several hundred nanometers, with lengths often extending tens of microns. Folded edge lines in the proposed folding structures in Figure 1 will produce line contrasts in TEM images similar to those seen for the "walls" of carbon nanotubes.[24]  A scanning TEM (STEM) image shows that central portions (the triple layer) of the fold exhibit higher intensities, indicating that it is thicker than the surrounding monolayer regions (Figure 2b). Residual particles preferentially decorate folding structures, implying that the particles are trapped inside the folding, or preferentially adhere around the folding edges during sample preparation as expected for the enhanced chemical reactivity of highly-curved graphene.

To confirm that the ribbon-like structures of Figures 2b are indeed a local triple-layer fold, we perform diffraction scanning across the folding structures using a nano parallel electron beam (NPEB). While scanning NPEB across the folding from locations 1 to 7 (Fig. 2b), we acquire diffraction patterns and monitor the transition across the fold. In location 1, which is outside of the fold, we observe one set of hexagonal diffraction spots (A spots) as shown in Fig. 2c. By comparing the (0-110) and (1-210) spot intensity, we confirm that this region is composed of monolayer graphene.[10,25] As we move the electron beam onto the fold structure, we find that another set of hexagonal spots (B spots) appears in the diffraction pattern, originating from the middle layer of folded graphene (Fig. 2d). The intensity of A spots inside the fold structure is twice that seen in monolayer regions, and a closer look at the (0-220) A spot reveals two distinct diffraction spots (0.4 degree rotation mismatch). This indicates that the diffraction pattern is comprised of three sets of hexagonal spots, and confirms that the folded region is a triple layer of graphene. We also note that the small angle splitting in the A spots is consistent with our observation that the two folding lines appear to be parallel in the STEM image. Figure 2f shows the variation of the diffraction spot intensity across the fold from locations 1 to 7. We observe that the intensity sum of A and B spots is a factor of three larger in folded regions than in monolayer regions, again consistent with our picture of a local triple-layer.

We can also easily identify grafolds with scanning electron microscopy (SEM) and atomic force microscopy (AFM). To do so, we transfer CVD graphene to a silicon oxide/silicon substrate. In the SEM image of Fig. 3a, we see dark line features across the graphene sample. The low-magnification STEM image of suspended CVD graphene shows that grafolds exhibit similar line features (Fig. 3b), which suggests that the dark lines in Fig. 3a are folds in graphene samples. When we perform the AFM scan on such dark lines, we observe a very similar ribbon-like feature as previously observed in the TEM study (Fig. 3c). Moreover, the sample indeed shows the higher height profile (~ 1 nm) inside the line (Fig. 3d). All these observations confirm that SEM can be used to identify folds in graphene. We also encounter more complex fold structures in graphene. Fig. 3e shows a quadruple fold structure in CVD graphene and the height difference (~ 0.7 nm) of observed steps is consistent with the thickness of double layer graphene.

Fig. 4a shows an atomic resolution TEM image of a pleat folding structure with parallel folding edge lines ($\theta = 27^o$). Inside the folding lines, graphene exhibits a moiré pattern, strong evidence of rotation-mismatched multilayer graphene.[11] From the fast Fourier transform (FFT) shown in Fig. 4b, we can identify two distinct hexagonal patterns, indicating that the middle graphene layer is rotated relative to aligned top and bottom layers. The observed folding angle of $\theta = 27^o$ is consistent with the angle calculated from the FFT of the micrograph, as the two sets of hexagonal patterns in Fig. 4b are mirror images from each other along the folding line (blue dotted line). We have also observed double folding structures with various folding angles including $\theta = 0^o$ (folding along the zigzag direction in graphene lattice).

The electrical and optical properties of graphene can be dramatically modified by changing the number and stacking relations of adjacent layers due to inter-layer interactions.[26-35] AB-stacked (Bernal stacked) bilayer and triple-layer graphene have distinct properties from monolayer graphene.[26-32] Non-regular stacking such as rotated two-layer graphene leads to interesting properties such as variable van Hove singularities and renormalization of charge carrier velocity.[33-35] Interesting questions arise upon laterally (in-plane) changing the number of graphene layers; in such situations, charge carriers will experience abrupt band-structure transformations at sites of layer number transitions. Even though there are a few previous theoretical calculations concerning electronic band structures of folded graphene, they have mainly focused on the bilayer graphene systems with single folds[36-38]; complex multiple fold structures with inhomogeneous layer numbers, such as pleat folds, have remained unexamined.

To better understand how folding affects graphene's material properties, we model the pleat folding structure in the supercell approximation.[39] The triple layer region is connected to the outer monolayer graphene by curved graphene regions formed from partial nanotubes. This simple model is then relaxed using the SIESTA code with a double-zeta (DZ) basis set until the forces are smaller than 0.04 eV/Angstrom.   The relaxed pleat folding structure ($\theta = 0^o$) originally formed from (8,8) and (13,13) armchair nanotubes is shown in Fig. 5a.   The length of the unit cell is 8.04 nm in the direction perpendicular to the fold and in the graphene plane, and 2.5 nm perpendicular to the graphene plane.

The geometry of the fold in the relaxed structure is influenced by the interlayer interactions and strain due to the curvature of the fold.   In Fig. 5a, it is apparent that the relaxed structure has flat regions which transition smoothly to bulges at the edges of the fold.   These bulges are similar to those found in collapsed nanotubes.[40]   The relaxed structure shows a stacking very similar to the ABC stacking pattern and the width of the folding can be indexed as $n = m = 14\frac{2}{3}$ (see the Appendix for the detailed indexing scheme).   However, unlike graphite, the grafold structure has regions of large curvature which necessitates that the layers are no longer parallel to each other.   Thus, the registry is changed as the graphene layer gets close to the fold, a result seen in previous studies of the bilayer graphene edge.[13]

We calculate the band structure of the grafold system along the direction of the fold as shown in Fig. 5b.   For a fold along the zigzag direction, the K point in the graphene Brillouin zone maps to $(1/3)*(2\pi/a)$ in grafold, which we label as "K" in the figure.   The band structure exhibits a semi-metallic character with the pi-states having a small, indirect overlap of ~8 meV near the "K" point.   This unconventional electronic band-structure is one example of what can be obtained in grafold nanostructures where many effects are simultaneously in play, such as curvature, interlayer coupling, layer registry, and confinement.  Detailed discussions on the effects of the layer registry can be found in the Appendix.  With narrow grafold structures, we also find that the occupied states in the vicinity of the Fermi level are localized near the folded region.  This can be seen in Fig. 5c where an isosurface of the local density of states, integrated over occupied states within 0.1 eV of $E_F$, is shown.  This localization may be useful for constructing electronic devices based on grafold nanostructures.

The formation of pleat folds occurs spontaneously during chemical vapor deposition (CVD) graphene synthesis or transfer processes.  Local wrinkles (or buckling) in graphene naturally form to relieve strain during the cool-down from high-temperature CVD growth temperatures,[41,42] during which the synthesis substrate contracts and the graphene slightly expands.[41-43]  These wrinkle structures often collapse to form folded structures when the copper substrate is etched away and the graphene is transferred to target substrates.  The TEM analysis of the angle dependent folding frequency and the folded area (fold frequency weighted by folding width) shows that folds have a fairly random distribution of orientations (Fig. 6a and 6b).

Grafold applications necessitate the ability to control the directions and locations of grafolds. We demonstrate that the directions and placement of folding formations can in fact be controlled by intentionally inducing, and then reconfiguring, surface curvatures in graphene during either the synthesis or transfer process.  Fig. 6c shows a route to controlled graphene folding in which we control the preferential folding directions by growing the graphene on a patterned substrate prior to transfer. In this method, a copper substrate with etched line trenches is used for CVD graphene growth. After growing the graphene, removing the copper substrate, and transferring the graphene to (isotropic) target substrates, the graphene exhibits folds more frequently along the original patterned copper trench direction. Graphene synthesized using this method and transferred to a Quantifoil holey carbon TEM grid is shown in Fig. 6d and 6e. Here, $\varphi = 0^o$ is the patterned trench direction in the copper substrate used for graphene synthesis. The histogram of angle-dependent folding frequency shows that the predominant folding formations occur along the trench direction (Fig. 6f and 6g) where a sample area of over 5,000 um$^2$ was investigated.

We propose another possible route for controlled placement of grafolds. Figure 7a shows a schematic of this proposal. After a standard CVD synthesis, graphene is transferred onto a substrate which has patterned metal (for example, Cu) rails or other removable features. The features are then etched away, allowing the local slack in the graphene to collapse and resulting in folding of the graphene at the desired regions. SEM images of graphene on a silicon substrate with a patterned copper rail, before and after the copper etching and graphene folding, are shown in Figures 7b and 7c, respectively; the observed features and apparent absence of graphene tearing suggest that the new slack results in graphene folding at the desired patterned locations.

Grafolds can also serve as a localized intercalation platform. To demonstrate this possibility, we intercalate $C_{60}$ into the grafolds. $C_{60}$ has been intercalated into graphite and various nanotubes in prior studies[44-47]. The introduction of $C_{60}$ could potentially induce interesting changes in the system's properties such as electronic band structure modifications.[45,46] In Fig. 8a, we show a TEM image of $C_{60}$ intercalation into a graphene edge fold. Even though the $C_{60}$ are randomly distributed away from the folded edge, we find that the $C_{60}$ are closely packed forming a single chain of $C_{60}$ molecules along the straight folded edge, confirming successful intercalation into the grafold. The available intercalants are essentially unlimited, opening up the possibility of tailoring grafold structures into a wide new class of materials with novel physical properties.

**Acknowledgements**

This research was supported in part by the Director, Office of Energy Research, Materials Sciences and Engineering Division, of the US Department of Energy under contract No. DE-AC02-05CH11231 which provided for preliminary TEM, SEM, and AFM characterization and partial theoretical support; by the National Science Foundation within the Center of Integrated Nanomechanical Systems, under Grant EEC-0832819, which provided for CVD graphene synthesis; and by the National Science Foundation under grant #0906539 which provided for design of the experiment, suspended sample preparation, and analysis of the results. Portions of the present study were performed at the National Center for Electron Microscopy, Lawrence Berkeley National Laboratory, which is supported by the U.S. Department of Energy under Contract # DE-AC02-05CH11231. Computational work was supported by National Science Foundation Grant No. DMR10-1006184 and computational resources have been provided by the Lawrencium computational cluster resource provided by the IT Division at the Lawrence Berkeley National Laboratory. K.K. acknowledges further support from a Samsung Scholarship, B.A. acknowledges support from the UC Berkeley A.J. Macchi Fellowship Fund in the Physical Sciences, and W.R. acknowledges support through a National Science Foundation Graduate Research Fellowship.

**Appendix A: Indexing of a pleat folding structure**

For a pleat folding in graphene, a translational vector between the two folding lines can be assigned as $\vec{w} = n\vec{a}_1 + m\vec{a}_2$. From this vector, the width $w = a\sqrt{n^2 + nm + m^2}$ and the folding angle $\theta = \tan^{-1}\left((n-m)/\sqrt{3}(n+m)\right)$ can be determined. However, finite local curvatures at the folded edges (Fig. 9a) may lead to ambiguous assignment of folding lines. Here, we suggest a way to define the folding edge lines to avoid this problem. Figure 9 show the schematic of this process. First, we define an axis of interest (blue dashed line) in the flat folded region around which the area has a certain well-defined stacking relation as shown in Fig. 9a. Any axis of interest chosen inside the flat region can assign unique folding lines (and indexing) for a pleat structure. There are three points of interest which are intersected by the blue dashed line in the three stacked graphene layers. Fig. 9b shows the top-view of the folding structure shown in Figure 9a. For the structure shown here, an ABC-stacking relation can be identified in the flat region. The points of interest are marked by the blue dot. Next, we unfold the pleat folding structure to obtain the corresponding flat graphene sheet while tracking the three points of interest as shown in Fig. 9c. We can now easily calculate the translational vectors between these three points. The folding edge lines (blue dashed lines in Fig. 9c) can be defined as the perpendicular bisectors of the lines connecting the three points of interest. In the pleat folding strcture shown in Fig. 9 (also the folding structure in Fig. 5a), the indexing is $n = m = 14\frac{2}{3}$. For the case of foldings in the zigzag or armchair directions of graphene lattice ($\theta = 0^o$ or $\theta = 30^o$), the indexing satisfies $n = m$ or $m = 0$, respectively.

**Appendix B: Band structures of shifted bilayer graphene**

We perform first-principles calculations for two layers of graphene with different relative shifts along the armchair direction to consider the effects that layer registry has on the electronic structure of the fold (Fig. 10). We use the plane-wave pseudopotential method[48] and density functional theory with the local density approximation[18,19], as implemented in the Quantum-ESPRESSO code.[49] The core-valence interaction is modeled using norm-conserving pseudopotentials.[20] A plane-wave cutoff of 60 Ry is used for the valence wavefunctions. The graphene lattice constant is 2.46 Å and the separation between the layers is fixed at 3.35 Å. The unit cell length perpendicular to the graphene sheets is 15 Å. The Brillouin zone is sampled with a $32 \times 32 \times 1$ Monkhorst-Pack k-point grid, and a Gaussian smearing of 0.05 eV is used for the electronic occupations.

As expected, the shift corresponding to AB stacking (Fig. 10c) between layers is most energetically favorable. Other configurations a, b, and d in Fig. 10 are higher in energy by 14, 6, and 1 meV/atom, respectively. The band structures for these configurations are shown in Fig. 10, Top. In agreement with previous studies,[27,50] the bands near $E_F$ for AB stacking show quadratic dispersion and a small band overlap very close to the K point. For AA stacking, linear bands cross at $E_F$ away from the K point. For a shift intermediate between AA and AB stacking (Fig. 10b), the conduction and valence bands show gaps at k-vectors away from the K point along the K-M and K-Γ directions. For another shift (Fig. 10d), we find two bands that cross away from K, above $E_F$, while there is a gap at K and along the K-Γ direction.

These calculations show that a shift in registry (relative translational position) of two graphene layers can create multiple valence band maxima and conduction band minima near, but shifted away from, the K point. The k-vector positions of these extrema relative to the K point and the band gaps are similar to what we calculate for the grafold structure. A similar effect was found in previous calculations for double-wall carbon nanotubes with shifted rotational orientations.[51] The indirect band overlap seen in the grafold band structure is not seen for two graphene layers, but could come about by other perturbations such as the curvature of the fold.

Although registry shifts away from AB stacking (or ABA or ABC for trilayer graphene) would not be expected for stacked graphene layers, the curved regions in grafold may allow the registry to change along the structure. Indeed, in our calculated grafold structure, the registry is not fixed throughout the fold. Therefore, these shifted registry effects can be present in the grafold band structure.

**Figures**

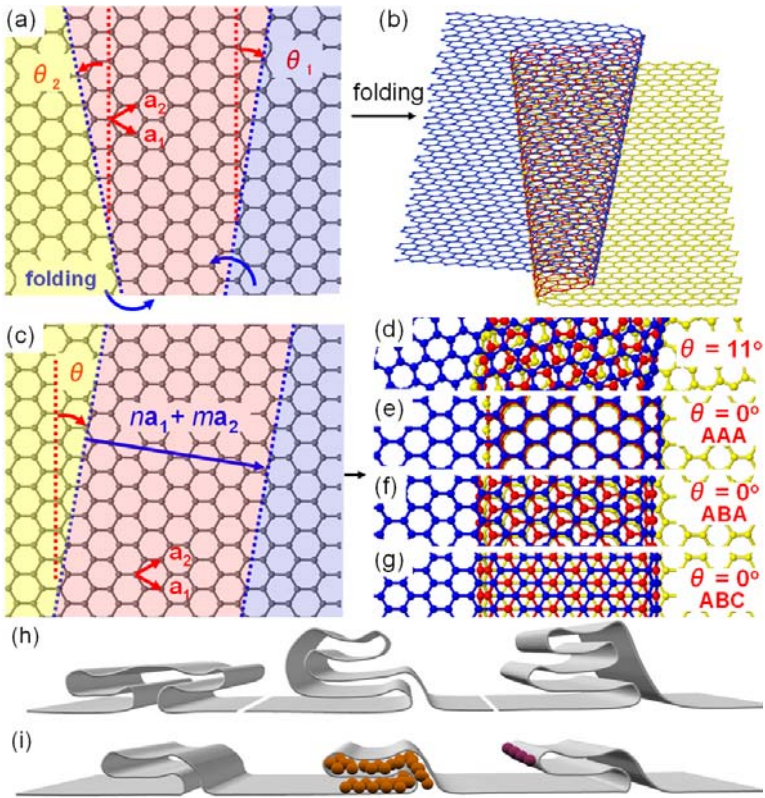

**FIG. 1. (Color online) Various multiple folding structures of graphene.** (a) Planar figure of graphene before double-folding. The dotted blue lines are folding lines, whose directions are measured relative to the direction of the $\vec{a}_1 - \vec{a}_2$ vector in the graphene lattice. (b) Double-folding of single layer graphene produces local triple-layer graphene. Upper, middle, and bottom graphene layers are shown in blue, red, and yellow, respectively. (c) A planar view of graphene before parallel double-folding (pleat folding). The translational vector between the two folding lines can be indexed as $n\vec{a}_1 + m\vec{a}_2$. (d) Top view of a folding structure with $\theta = 11^o$. Top and bottom graphene layers have the same rotational direction. (e-g) Top views of folding structures with $\theta = 0^o$, where foldings occur along the zigzag lattice direction of graphene. The layer stacking can vary depending on where the folding lines occur. (e) AAA stacking of the folding structure with $n = m = 4\frac{1}{2}$. (f) ABA stacking with $n = m = 4\frac{1}{2}$. (g) ABC stacking with $n = m = 4\frac{2}{3}$. (h) Various complex folding structures with quadruple-foldings, including a box-pleat (left). (i) Periodic double folding superstructure. Folds can be used as a local intercalation platform, and highly curved edges can be functionalized with foreign materials.

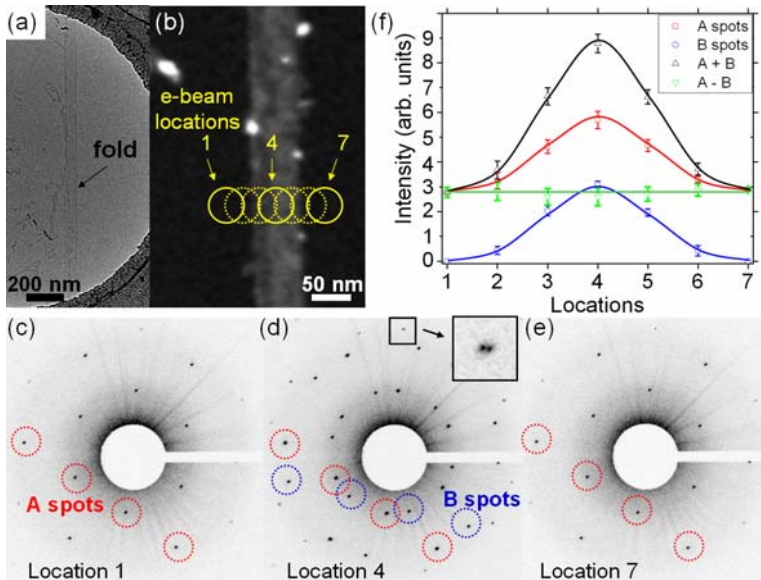

**FIG. 2. (Color online) TEM and diffraction analysis of double folding structures.** (a) TEM image of a ribbon-like folding structure in suspended graphene. (b) HAADF (High-angle annular dark field) STEM image of a graphene sheet with a folding structure. Circles denote different locations 1-7 for diffraction acquisition using a nano-parallel electron beam (NPEB). The NPEP has a diameter of 45 nm. (c-e) Series of diffraction patterns across the folding structure. (c) Diffraction pattern at location 1. (d) Diffraction pattern at location 4, at the folding. Inset shows a zoom-in image of the (0-220) A spot, which shows the splitting of two spots with a very small rotation angle of $0.4^o$. (e) Diffraction pattern at the location 7. (f) Diffraction spot intensity (A spot, B spot, their sum, and their difference) variations at different locations.

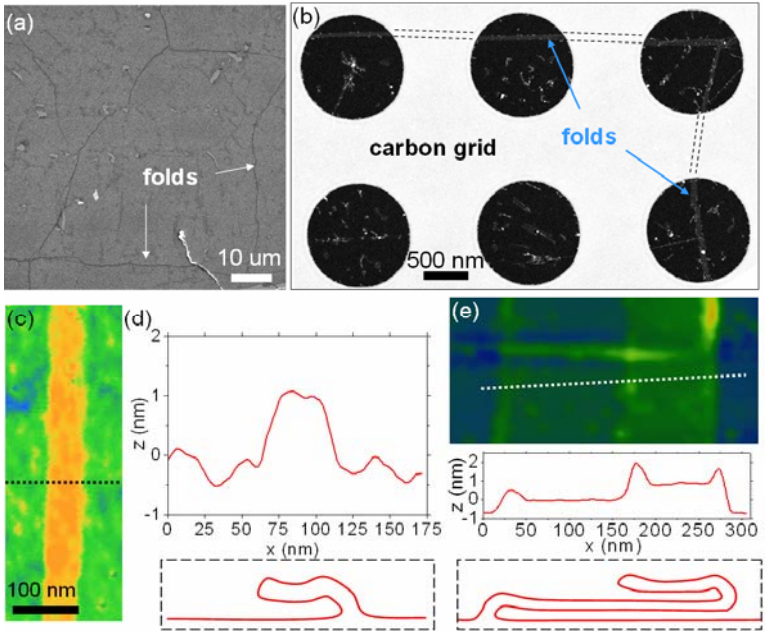

**FIG. 3. (Color online) SEM, AFM, and STEM images of folds in CVD graphene.** (a) SEM images of CVD graphene transfered to a silicon oxide/silicon substrate. Folds appear dark relative to monolayer graphene. (b) HAADF STEM image of CVD graphene transfered to a Quantifoil holey carbon TEM grid.   The white background is the region of the amorphous carbon film and the dashed lines are a guide to the eyes for folds in graphene. (c) AFM scan of graphene fold in CVD graphene transfered to silicon oxide/silicon substrate. (d) Top - The height profile of the folding structure along the dotted line in Fig c. Bottom - Proposed folding structure (double folding) from the AFM data. (e) Top - AFM scan of multiple folding structure in graphene. The white dotted line is used for the height profile scan. Middle - The height profile of the folding structure. The height difference (0.7 nm) of observed steps is consistent with the thickness of double layer graphene. Bottom - Proposed folding structure (quadruple folding) from the AFM data.

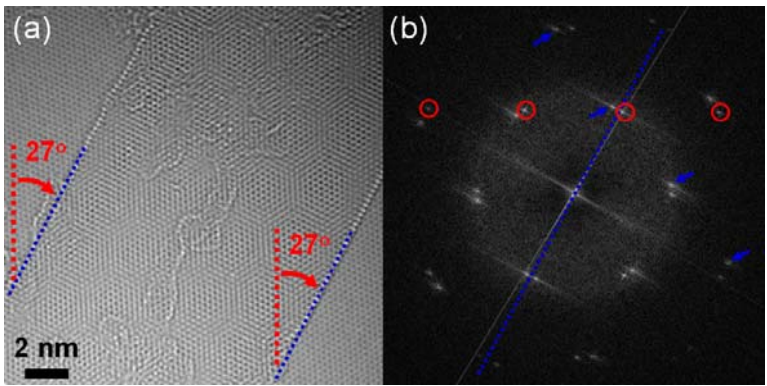

**FIG. 4. (Color online) Atomic resolution TEM image of a pleat folding structure.** (a) A double folding structure with parallel folding lines with $\theta = 27^o$. To the left and right of the folding structure is monolayer graphene. The folding edges are overlaid with blue dotted lines. (b) Fast Fourier transform (FFT) of Figure 4a, which shows two sets of hexagonal patterns. The circled set of hexagonal pattern (red) is from middle layers. The blue dotted line is drawn to be parallel to the folding lines. The hexagonal set marked by blue arrows is from top and bottom graphene layers, which have the same lattice directions after folding.

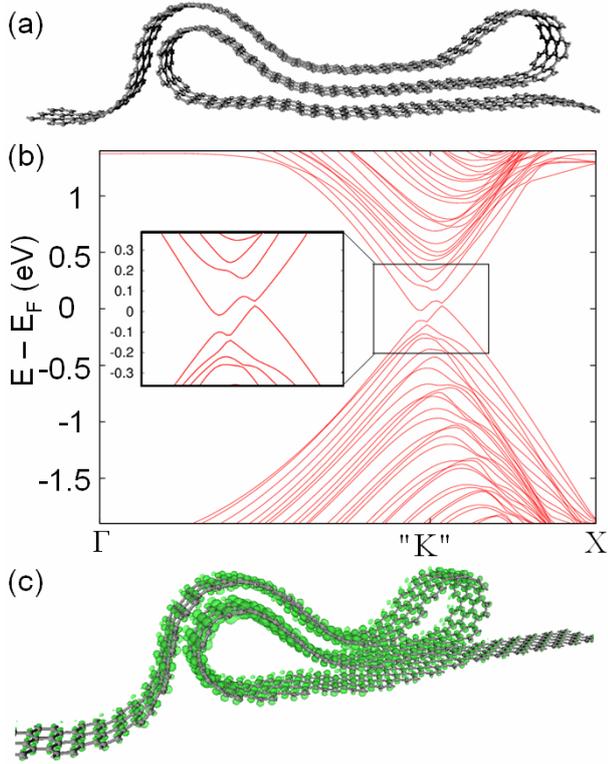

**FIG. 5. (Color online) Theoretical calculation of double folding structures in graphene.** (a) Model of the pleat folding structure of $\theta = 0^o$ that has been relaxed. The fold is periodic into the page, although only a finite width is shown here for clarity. (b) Electronic band structure of the pleat folding along the direction of the fold. The system is semi-metallic with a slight indirect overlap between the occupied and unoccupied states. (c) Calculated local density of states integrated for the occupied states from 0.1 eV below $E_F$ to $E_F$ for a narrow double folding with a triple layer region of length ~ 1 nm. The shown iso-surface is 10% of the maximum value and shows the occupied states near the Fermi level are localized to the region near the fold.

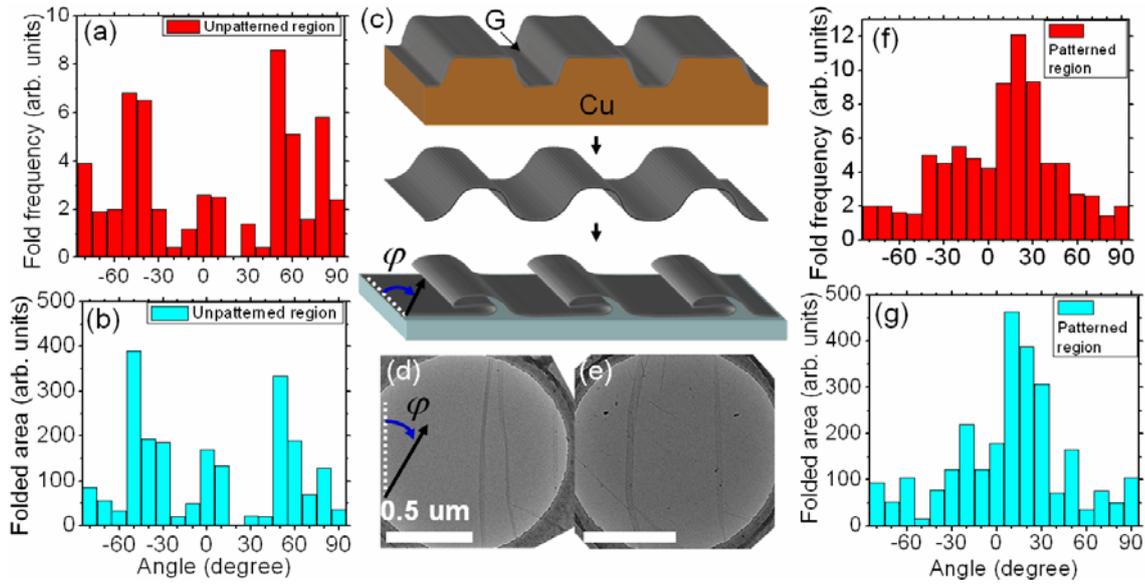

**FIG. 6. (Color online) Controlled folding formations in graphene: comparison of graphene folding formations from regular and patterned copper substrates.** (a) Histogram of angle-dependent fold formation frequency in graphene grown on unpatterned copper foil. (b) Histogram of angle-dependent folded area in folds in graphene grown on unpatterned regular copper foil. (c) Schematic process of directional control of folding formations in graphene during synthesis. A copper substrate with patterned etched trenches is used for CVD graphene growth. After transferring the graphene to other substrates, foldings are more frequently aligned with the patterned direction. (d-e) Graphene synthesized using the patterned copper substrate and transferred to a Quantifoil holey carbon TEM grids. Aligned folding formations are observed. $\varphi = 0^o$ is the pattern direction in the copper substrate. (f-g) Histogram of angle-dependent fold formation frequency and folded area in graphene grown on etched line trenches of copper foil. The angle at zero degrees is the direction of the patterns.

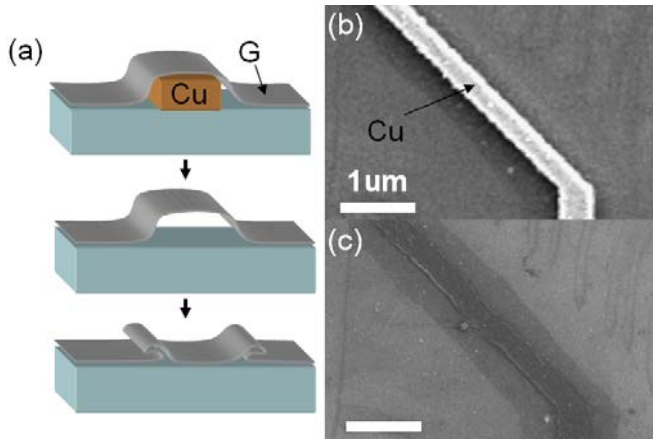

**FIG. 7. (Color online) Proposal for controlled placement of grafold.** (a) Schematic process of controlled placement of folds in graphene. Graphene is transferred onto a substrate which has patterned metal features. After underetching of the metal features, the graphene collapses and may fold along these pre-patterned regions. (b) SEM image of graphene draped over a copper metal pattern on a silicon substrate. (c) SEM image of the same region after the copper is underetched. The image is suggestive of graphene folding aligned with the removed copper structure.

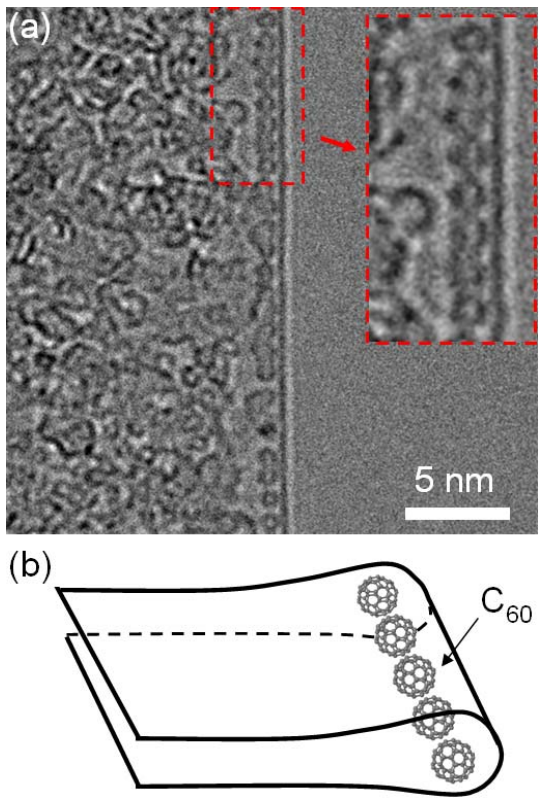

**FIG. 8. (Color online) Intercalation of $C_{60}$ into a single fold in graphene.** (a) TEM image of a graphene edge fold with intercalated $C_{60}$. Along the straight fold edge, $C_{60}$ are closely packed forming a single chain of $C_{60}$. (b) Schematic view of the fold structure with local $C_{60}$ intercalation.

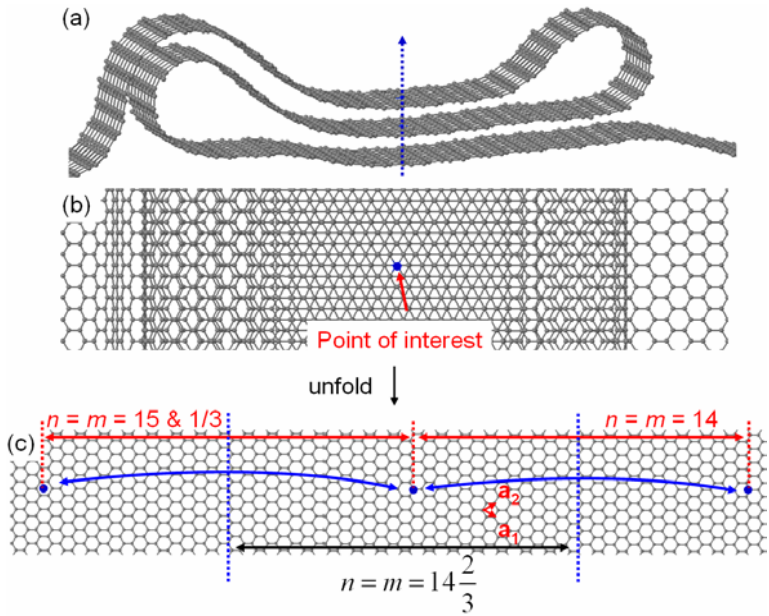

**FIG. 9. (Color online) Schematic of indexing of a pleat folding structure.** (a) Assignment of an axis of interest in the folding structure. There are three points of interest which are intersected by the blue dashed line in the three stacked graphene layers. (b) Top-view of the folding structure shown in Figure 9a. An ABC-stacking relation can be identified in the flat region and the points of interest are marked by the blue dot. (c) Unfolding the structure to obtain a flat graphene sheet and finding the translational vectors connecting the three points of interest to index the folding lines (blue dashed lines). In the folding structure shown in the schematic, the indexing is $n = m = 14\frac{2}{3}$.

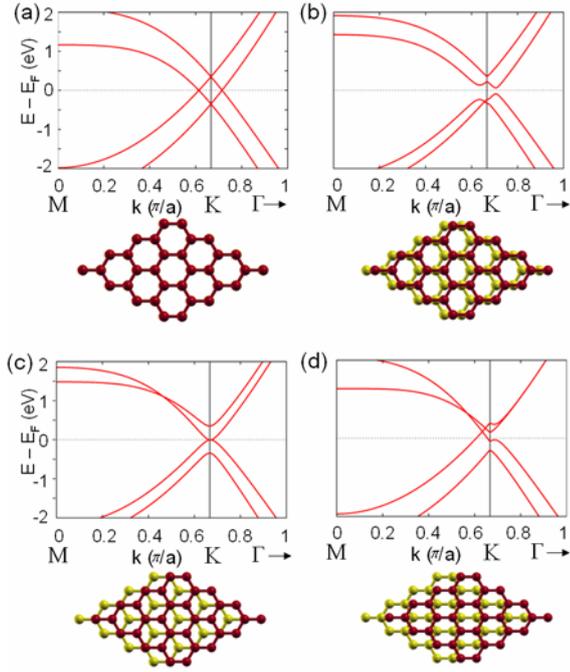

**FIG. 10. (Color online) Calculated band structures of shifted bilayer graphene.** (a)-(d), Calculated band structures along the M→K→Γ direction for two layers of graphene (Top) with corresponding atomic structures (Bottom) having different relative shifts in the armchair direction. (a) and (c) are for AA and AB stacking, respectively, while (b) and (d) are intermediate stackings. In (a), two pairs of bands cross at $E_F$ away from the K point and have linear dispersion. In (c), conduction and valence bands touch at $E_F$ at the K point and have quadratic dispersion. The intermediate case (b) shows a gap, with band extrema away from the K point. In (d), a band crosses $E_F$ at two points away from the K point.